\documentclass[aps,10pt,prb,twocolumn,letterpaper,superscriptaddress]{revtex4-2}

\usepackage{graphicx}
\usepackage{dcolumn}
\usepackage{bm}

\begin{document}

\preprint{APS/123-QED}

\title{Quantum emitter formation dynamics and probing of radiation induced atomic disorder in silicon}

\author{Wei Liu$^\dagger$}
\email{weiliu01@lbl.gov }
\thanks{$^\dagger$These authors contributed equally }
 \affiliation{ Accelerator Technology and Applied Physics Division, Lawrence Berkeley National Laboratory, Berkeley, CA 94720, USA }
 
\author{Vsevolod Ivanov$^\dagger$ }\affiliation{ Accelerator Technology and Applied Physics Division, Lawrence Berkeley National Laboratory, Berkeley, CA 94720, USA }\affiliation{Molecular Foundry, Lawrence Berkeley National Laboratory, Berkeley, CA 94720, USA}
\author{Kaushalya Jhuria}\author{Qing Ji}\author{Arun Persaud} 
 \affiliation{ Accelerator Technology and Applied Physics Division, Lawrence Berkeley National Laboratory, Berkeley, CA 94720, USA }
\author{Walid Redjem} \affiliation{Department of Electrical Engineering and Computer Sciences, University of California, Berkeley, CA 94720, USA}

\author{Jacopo Simoni}%
\affiliation{Molecular Foundry, Lawrence Berkeley National Laboratory, Berkeley, CA 94720, USA}%

\author{Yertay Zhiyenbayev}
 \affiliation{Department of Electrical Engineering and Computer Sciences, University of California, Berkeley, CA 94720, USA}

\author{Boubacar Kante}\affiliation{Department of Electrical Engineering and Computer Sciences, University of California, Berkeley, CA 94720, USA}
\author{Javier Garcia Lopez}
 \affiliation{Centro Nacional de Aceleradores (U. Sevilla, CSIC, J. de Andalucia), Seville, Spain, 41092, Seville, Spain}
\author{Liang Z Tan}\affiliation{Molecular Foundry, Lawrence Berkeley National Laboratory, Berkeley, CA 94720, USA}%
\author{Thomas Schenkel}
\affiliation{ Accelerator Technology and Applied Physics Division, Lawrence Berkeley National Laboratory, Berkeley, CA 94720, USA }


\begin{abstract}
Near infrared color centers in silicon are emerging candidates for on-chip integrated quantum emitters, optical access quantum memories and sensing. We access ensemble G color center formation dynamics and radiation-induced atomic disorder in silicon for a series of MeV proton flux conditions. Photoluminescence results reveal that the G-centers are formed more efficiently by pulsed proton irradiation than continuous wave proton irradiation. The enhanced transient excitations and dynamic annealing within nanoseconds allows optimizing the ratio of G-center formation to nonradiative defect accumulation. The G-centers preserve narrow linewidths of about 0.1 nm when they are generated by moderate pulsed proton fluences, while the linewidth broadens significantly as the pulsed proton fluence increases.  This implies vacancy/interstitial clustering by overlapping collision cascades. Tracking G-center properties for a series of irradiation conditions enables sensitive probing of atomic disorder, serving as a complimentary analytical method for sensing damage accumulation. Aided by \textit{ab initio} electronic structure calculations, we provide insight into the atomic disorder-induced inhomogeneous broadening by introducing vacancies and silicon interstitials in the vicinity of a G-center. A vacancy leads to a tensile strain and can result in either a redshift or blueshift of the G-center emission, depending on its position relative to the G-center. Meanwhile, Si interstitials lead to compressive strain, which results in a monotonic redshift. High flux and tunable ion pulses enable the exploration of fundamental dynamics of radiation-induced defects as well as methods for defect engineering and qubit synthesis for quantum information processing. 
\end{abstract}

\maketitle


\section{ Introduction }

 Programmable quantum information processing and unbreakable secure quantum networks linked by trusted repeater nodes require the robust generation and manipulation of single and few quanta in a scalable platform with long coherence time \cite{moody20222022,gimeno2019deterministic,gao2012observation}. On-chip integrated quantum photonic circuits in silicon have been used to demonstrate the implementation of quantum protocols via multiplexed linear optical elements and nonlinear electro-optic effects, thanks to its mature manufacturing and low optical losses \cite{wang2018multidimensional,wang2020integrated}. However, because of the indirect band gap of silicon, it has been a long-standing challenge to fabricate highly efficient silicon-based quantum emitters enabling on-demand single photons, entangled-photon pairs, spin-photon entanglements, squeezed light, etc. Though integration of heterogeneous active sources has been implemented using \cite{fathpour2015emerging}, e.g., III–V quantum dots, color centers in diamond or SiC, and spontaneous parametric down-conversion (SPDC) nonlinear materials, searching for emissive qubits in Si as a host material \cite{davies1989optical} is desirable for the CMOS process compatibility and scale up purposes. Alternatively, direct formation of telecom-band light emitting defects in silicon (Si) through non-ionizing energy loss (NIEL) of energetic particles has become an emerging approach for on-chip integration, as recently demonstrated using carbon-based defect complexes of G- and T-centers \cite{higginbottom2022optical, redjem2020single,schenkel2022exploration,komza2022indistinguishable,redjem2023all}, as well as rare-earth erbium ions \cite{gritsch2022narrow,yin2013optical}. On the other hand, it is well-known that irradiating a semiconductor with  energetic particles gives rise to structural damage and defect formation via NIEL processes \cite{kraner1982radiation}.  This leads to the degradation of device performance, e. g. in radiation detectors operating in (high) radiation environments.  Radiation damage effects include reduced sensitivity of photodiodes \cite{moll2018displacement}, decreased efficiency of light emitting diodes (LED) and lasing threshold shifts \cite{polyakov2013radiation}, as well as degrading reliability of power electronics \cite{pearton2021opportunities}. The build-up of stable radiation damage often proceeds via complex dynamic annealing (DA) processes, involving point defect migration and interaction, which often occurs during the ion implantation/irradiation step for actual devices fabrication, and depends on the density of collision cascades and the ion beam flux \cite{nordlund2018primary, wallace2015radiation}. Such dose-rate effects on radiation damage and defect interaction dynamics in Si have been studied by various analytical methods, including deep-level transient spectroscopy (DLTS) \cite{hallen1991influence,leveque2002dose}, charge collection efficiency (CCE) measurements \cite{borer2000charge}, Rutherford back scattering spectroscopy (RBS) \cite{titov1996defect, wallace2018deterministic}, X-ray diffraction analysis (XRD), and positron annihilation Doppler broadening spectroscopy (PAS) \cite{minagawa2017effect}. However, the effect of ion dose rate on the formation dynamics of color centers and the correlation between inhomogeneous broadening and atomic radiation disorder have rarely been studied, even though understanding these processes is essential for optimizing the formation yield, coherence, luminescence efficiency and deterministic positioning of these defect emitters in Si quantum photonic circuits \cite{redjem2023all}. 

In this study, we explore radiation-induced G-center formation dynamics in silicon wafers under various proton irradiation conditions: (1) by using the induction linear accelerator that is part of the Neutralized Drift Compression Experiment (NDCX-II) at Lawrence Berkeley National Laboratory, which delivers a $\sim 1$ MeV intense nanosecond-pulsed proton beam, and allows us access to transient radiation effects on defect dynamics far from equilibrium at the nanosecond time scale \cite{schenkel2013towards,schenkel2022exploration}; (2) by using 1 MeV continuous wave proton irradiation with orders of magnitude lower dose rates performed in \emph{Centro Nacional de Aceleradores}, Seville, Spain \cite{lopez2000cna}. Our present study complements earlier silicon irradiation studies with pulsed proton and ion pulses from laser-accelerators and higher damage rates \cite{redjem2022defect}. We compare the G-center optical properties characterized by time-resolved photoluminescence (PL) and reveal the dose rate effect on color center formation efficiency and optical linewidth.  Furthermore, we perform \textit{ab initio} electronic structure calculations, which provide insight into the atomic-disorder-induced inhomogeneous broadening, by introducing vacancies and Si interstitials in the vicinity of a G-center. We highlight the approach of using pulsed proton irradiation on the as-received carbon silicon wafer to generate color center complexes with high formation efficiency while preserving the narrow linewidth, and demonstrate using G-centers as a sensitive optical probe for atomic radiation disorder.

\section{SAMPLES AND MEASUREMENTS}
Float zone silicon with residual carbon, (111) crystallographic plan, $\sim$100 Ohm$\cdot$cm resistivity, was used as the starting material for proton beam irradiation. Note that the as-received carbon in the silicon samples is likely due to the extended storage and no additional carbon implantation or annealing was performed in the experiments we report here. Such a starting material circumvents the typical carbon implantation step for G-center formation, which introduces the radiation damage \cite{beaufils2018optical,berhanuddin2012co}. Secondary ion mass spectrometry (SIMS) prior to irradiation showed a carbon atom concentration of up to 10$^{20}$ C/cm$^{3}$ near the native oxide surface, which drops to the SIMS sensitivity limit of 10$^{16}$ C/cm$^{3}$ at a depth of 150 nm, corresponding to a carbon areal density of 2 $\times$ 10$^{14}$ C/cm$^{2}$ \cite{schenkel2022exploration}.

Pulsed proton irradiation was performed by NDCX-II, which delivers up to 1.1 MeV protons for a $\sim$10 ns pulsed length, with a repetition rate of 1 shot/45 s. The $\sim$1 MeV proton generates dilute vacancies and interstitials at a rate of $\sim$2 × 10$^{-4}$ damage events/(proton$\cdot$nm) in the top 1 µm of the samples \cite{schenkel2022exploration}. A 4 µm thick aluminum foil in front of the sample was used to filter the residual H$_{2}^{+}$, which results in a 0.9 MeV proton landing on a $\sim$2 mm diameter area of the sample. Two different dose rates of $7.9 \pm 1.6 \times 10^{10}$ protons/cm$^{2}$/pulse (transient ion flux of 7.9$\times$10$^{18}$ protons/cm$^{2}$/s) and $6.9 \pm 1.6 \times 10^{9}$ protons/cm$^{2}$/pulse (transient ion flux of 6.9$\times$10$^{17}$ protons/cm$^{2}$/s), determined by Faraday cup measurements on a series of pulses collected both before and after the NDCX-II was used to irradiation the sample. On the other hand, 1 MeV continuous wave proton irradiation was performed on the same batch of silicon wafers. For the fluences of 1$\times$10$^{11}$ cm$^{-2}$ and 1$\times$10$^{12}$ cm$^{-2}$, the beam current was $\sim$10 pA (ion flux of 8×10$^{9}$protons/cm$^{2}$/s), while for the higher fluences (1$\times$10$^{13}$ cm$^{-2}$ and 1$\times$10$^{14}$ cm$^{-2}$), the beam current was $\sim$1 nA (ion flux of 8$\times$10$^{11}$ protons/cm$^{2}$/s).

PL spectra were recorded at 4 K using a confocal near-infrared micro-PL setup. Optical excitation was performed with a pulsed laser (20 Mhz, 532 nm) focused onto the sample via an objective (N.A. = 0.85), which provides about 0.6 $\mu$m diameter at focus and 1 $\mu$m field of depth. The PL spectrum was collected by the same objective and guided to a spectrometer (with grating of 1200 grooves/mm) coupled to an InGaAs camera, which provides better than 0.03 nm spectral resolution. The time-resolved PL was measured by using a superconducting single-photon detector (SNSPD) with a band pass filter (tunable center wavelength around 1300 nm and 12 nm bandwidth), which provides a detection efficiency of 85\%.

\section{Experimental results}
\subsection{Proton fluence and dose rate effect on G-center formation dynamics}
\begin{figure}[ht]
\includegraphics[width=\columnwidth]{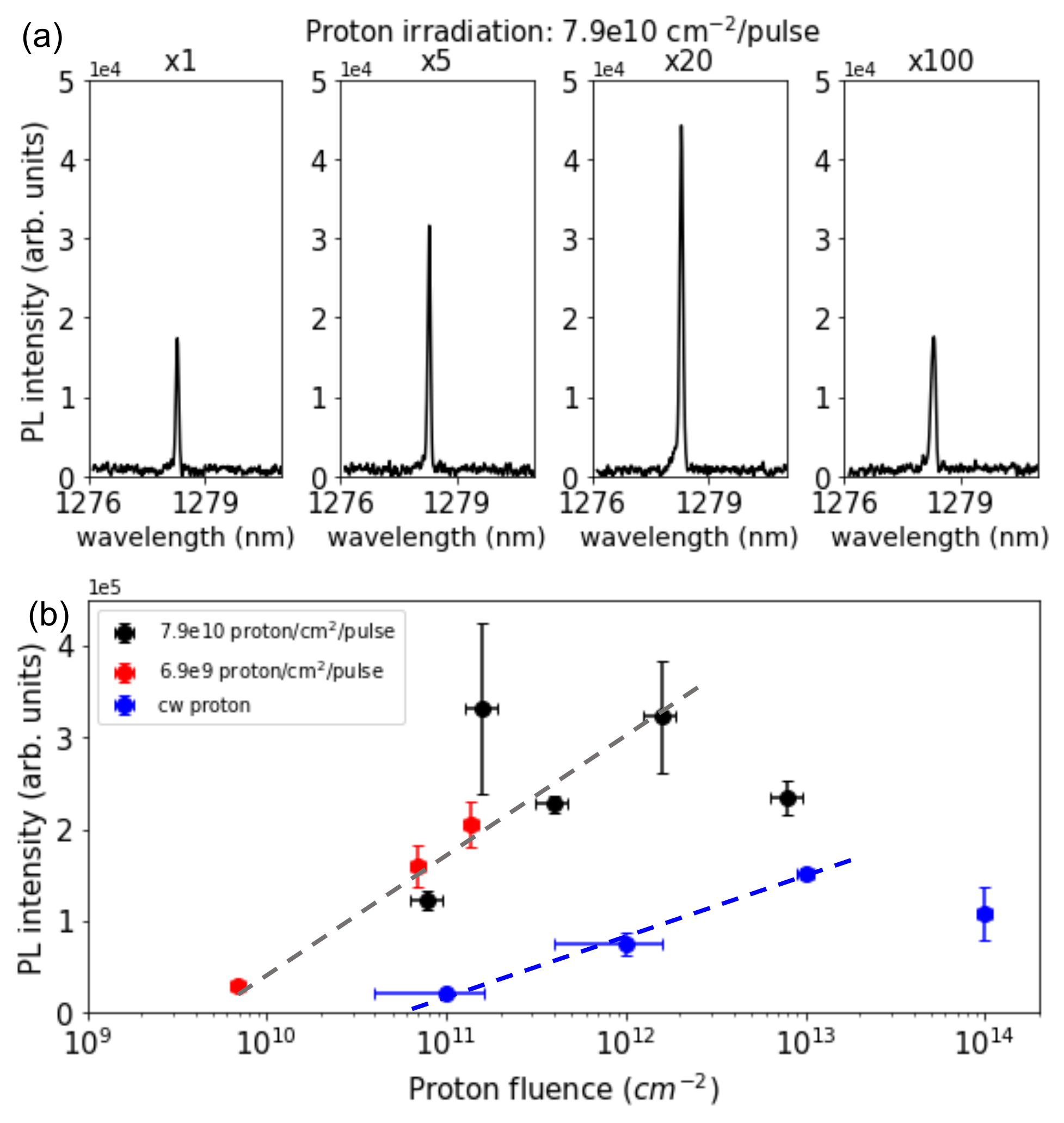}
\caption{(a) example of PL spectra of G-center created by 1 shot, 5 shots, 20 shots, and 100 shots of $7.9 \pm 1.6 \times 10^{10}$protons/cm$^{2}$/pulse, respectively, with pump power of laser at 0.3 mW; (b) comparison of integrated PL intensity of G-center generated by $7.9 \pm 1.6 \times 10^{10}$protons/cm$^{2}$/pulse and $6.9 \pm 1.6 \times 10^{9}$ protons/cm$^{2}$/pulse and cw proton, with varying proton fluence from 10$^{11}$ cm$^{-2}$ to 10$^{14}$ cm$^{-2}$.}
\label{fig1}
\end{figure}

Figure \ref{fig1}(a) presents an example of the ensemble G-center PL zero phonon line (ZPL) around 1278 nm generated by pulsed proton irradiation of 1 shot, 5 shots, 20 shots and 100 shots of $7.9 \pm 1.6 \times 10^{10}$cm$^{-2}$/pulse. The gradual increase of PL intensity of G-centers as the proton fluence up to $1.6 \times 10^{12}$ cm$^{-2}$ (for 20 shots) is evidence for the direct formation of G-centers by proton induced Si intersitials combing with carbon pairs formed by a substitutional carbon trapping a migrating carbon interstitial\cite{beaufils2018optical,berhanuddin2012co}. 
However, the PL intensity drops significantly as the proton fluence increases to $7.9 \pm 1.6 \times 10^{12}$cm$^{-2}$/pulse by 100 shots, which can be related to either an increase of the nonradiative channel by radiation damage, or a reduced yield of G-centers. Noted that the repeated pulses can lead to annealing or may alter the vacancies/interstitials and G-centers that had been formed by the preceding pulses. Here, we expect the heating effect from the energy deposition by the proton pulses to be marginal, considering only a $\sim 4$ K temperature variation generated by each pulsed proton at a depth of 0.5 µm in the silicon  \cite{schenkel2022exploration}.

We further compare the dose rate effect on the G-center formation dynamics. Figure \ref{fig1}(b) summarizes the integrated PL intensity of G-center generated by multiple pulsed proton shots with $7.9 \pm 1.6 \times 10^{10}$ cm$^{-2}$/pulse (black dot) and $6.9 \pm 1.6 \times 10^{9}$ cm$^{-2}$/pulse (red), and cw protons (blue dot), with varying proton fluence ranging from 10$^{11}$ cm$^{-2}$ to 10$^{14}$ cm$^{-2}$. The scaling of PL intensity with proton fluence is larger for pulsed irradiation with a power law of exponent of 0.65, compared to 0.17 for the cw case. This shows that pulsed irradiations with higher proton flux favor G-center formation. 

\subsection{Carrier recombination dynamics by time-resolved PL}
\begin{figure}[t]
\includegraphics[width=\columnwidth]{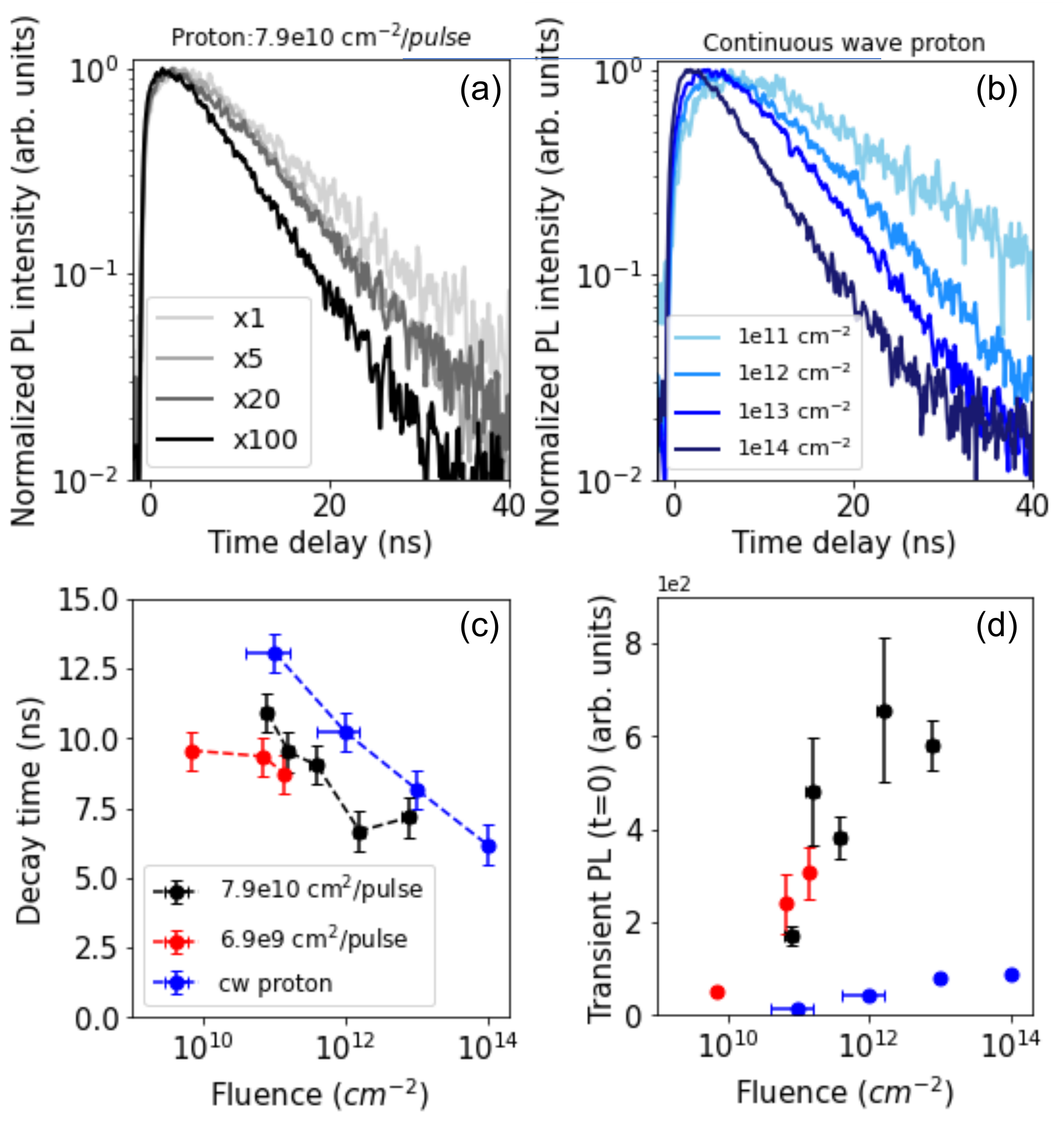}
\caption{(a) time-resolved PL of G-centers created by pulsed proton ($7.9 \pm 1.6 \times 10^{10}$ protons/cm$^{2}$/pulse) irradiation by varying the number of shots from 1 to 100, respectively; (b) time-resolved PL of G-centers created by cw proton irradiation with fluence varying from $1 \times 10^{11}$ cm$^{-2}$ to $1 \times 10^{14}$ cm$^{-2}$; (c) effective PL decay time extracted using single exponent fitting of the experiments as a function of proton fluence by using a pulse of $7.9 \pm 1.6 \times 10^{10}$ protons/cm$^{2}$/pulse (black dots), $6.9 \pm 1.6 \times 10^{9}$ protons/cm$^{2}$/pulse (red dots), and cw proton irradiation (blue dots and dashed-lines as a guide to the eyes); (d) the corresponding transient PL intensity at initial delay as a function of proton fluence by using pulsed and cw proton irradiation. }
\label{fig2}
\end{figure}

We perform time-resolved PL measurements to investigate the dose rate effect on the carrier recombination dynamics of radiation-induced G-centers. Figure \ref{fig2} (a) and (b) show the example of PL decays of G-centers created by pulsed proton ($7.9 \pm 1.6 \times 10 ^{10}$protons/cm$^{2}$/pulse) by varying the number of shots from 1 to 100, and by cw proton irradiation with fluence varying from $1 \times 10^{11}$ cm$^{-2}$ to $1 \times 10^{14}$ cm$^{-2}$. For both pulsed and cw irradiation, the PL decay becomes faster with increasing proton fluence, which is likely due to the increased density of radiation-damage-induced nonradiative defects, which leads to Shockley-Read-Hall (SRH) recombination \cite{berhanuddin2012co}. As detailed in the Appendix \ref{appendix:a}, the increase of PL decay time with laser pump power is consistent with the typical feature of the gradual saturation of the nonradiative SRH process. The nonradiative defects can be the vacancy-related deep level states, e.g. divacancy with 0.42 eV below the conduction band or higher order vacancy clusters \cite{leveque2002dose,huhtinen2002simulation}. 

Note that the radiative lifetime ($\tau_{r}$) for ensembles of G-centers is expected to remain constant within a relatively dilute density range (with no or only very minimal dipole-dipole coupling between G-centers) and is dominated by the intrinsic dipole moment of individual G-centers. Such assumptions are reasonable in our sample, in which the upper bound of G density within 50 nm is $1 \times 10^{12}$ cm$^{-2}$ created by the $1 \times 10^{14}$ cm$^{-2}$ proton irradiation, considering the $2 \times 10^{-4}$ damage events/(proton$\cdot$nm) generated by the $\sim$1 MeV proton. It is worth noting that, as shown in figure \ref{fig2} (b), the rising time of PL intensity ($\tau_{rise}$) to its maximum is gradually delayed as the cw proton fluence decreases. It indicates a longer carrier diffusion length before being captured by either the G-center or nonradiative defects with larger averaging distance in the dilute density range. As a comparison, the $\tau_{rise}$ of G-center is sharp in the case of pulsed irradiation, which can be ascribed to the fast capture of carriers by the higher amount of G-center formation or nonradiative defect densities, therefore leading to a short carrier diffusion length. 

Figure \ref{fig2}(c) summarizes the effective PL decay time of G-centers extracted by single exponent fitting of the experiments as a function of proton fluence, using multi-pulses of $7.9 \pm 1.6 \times 10^{10}$ protons/cm$^{2}$/pulse (black dots), $6.9 \pm 1.6 \times 10^{9}$ protons/cm$^{2}$/pulse (red dots), and cw proton irradiation (blue dots). As the cw proton fluence is increased from $1 \times 10^{11}$ cm$^{-2}$ to $1 \times 10^{14}$ cm$^{-2}$, the decay time  gradually decreases from 13 ns to 6 ns. The PL lifetime of G-centers formed by the pulse irradiation follows a similar trend as the one for cw. Referring to the reported single isolated G-center decay time of $>$ 45 ns \cite{redjem2020single}, the shorter decay time we observe here is composed of radiative channel of G-center and parasitic nonradiative defect channels. Assuming a $\tau_{r}$ = 45 ns, the quantum efficiency of ensemble G-centers is about 29\%-13\% in our case, corresponding to a nonradiative lifetime {$\tau_{nr}$} of 18 ns – 7 ns. On the other hand, the $\sim 6$ ns decay time at $1 \times 10^{14}$ cm$^{-2}$ proton fluence is consistent with the reported G-centers generated by similar proton irradiation \cite{beaufils2018optical}, which suggests a characteristic decay time dominated by the nonradiative defects induced by the elevated radiation damage. 

Figure \ref{fig2} (d) compares the corresponding transient PL intensity of G-centers at initial delay as a function of proton fluence by using pulsed and cw proton irradiation. Importantly, the transient PL intensity at initial delay can be safely assumed to be proportional to the G-center density $n_{G}$, namely $\propto n_{G}$/$\tau_{r}$, if the $\tau_{nr}$ is orders of magnitude larger than the time resolution \cite{liu2016exciton,rossbach2014high}, a condition which is fulfilled in our study considering the ns range of $\tau_{nr}$ compared to the 100 ps resolution of our setup. As presented in \ref{fig2} (d) the scaling of transient PL intensity ($\propto n_{G}$) to proton fluence is much larger for the pulsed irradiation compared to the cw case, which further demonstrates that the pulsed irradiation is more favorable for G-center formation. As an example, even when both beams have an equivalent 10$^{13}$ cm$^{-2}$ proton fluence, the transient PL intensity under pulsed irradiation is 6 times higher than that of cw one. On the other hand, a slightly shorter decay time (7 ns) under pulsed irradiation compared to that of 8.5 ns under cw at 10$^{13}$ cm$^{-2}$ proton fluence indicates that the pulse irradiation also introduces more nonradiative defects. Such a dose rate effect on G-center formation can be related to the protons generating enhanced transient excitations in the silicon, followed by the dynamical annealing by the instantaneous energy deposition. Such transient excitations by pulsed protons affect the mobility of interstitials/vacancies, vary the formation energy by transient Fermi level, and modulate the rate of G-center creation versus destruction, as well as the rate of vacancy/interstitial accumulation versus annealing within the nanosecond range \cite{wallace2017role}. In addition, our recent theoretical study indicates that optically active G-centers can be altered into an optically inactive configuration by overcoming a  conversion barrier of about 0.15 eV \cite{ivanov2022effect}. Such a perturbation of G-centers formed by previous protons can be triggered by the incoming protons under the cw irradiation. 

\subsection{Impact of dose rate effect on G-center linewidth broadening}
\begin{figure}[h]
\includegraphics[width=\columnwidth]{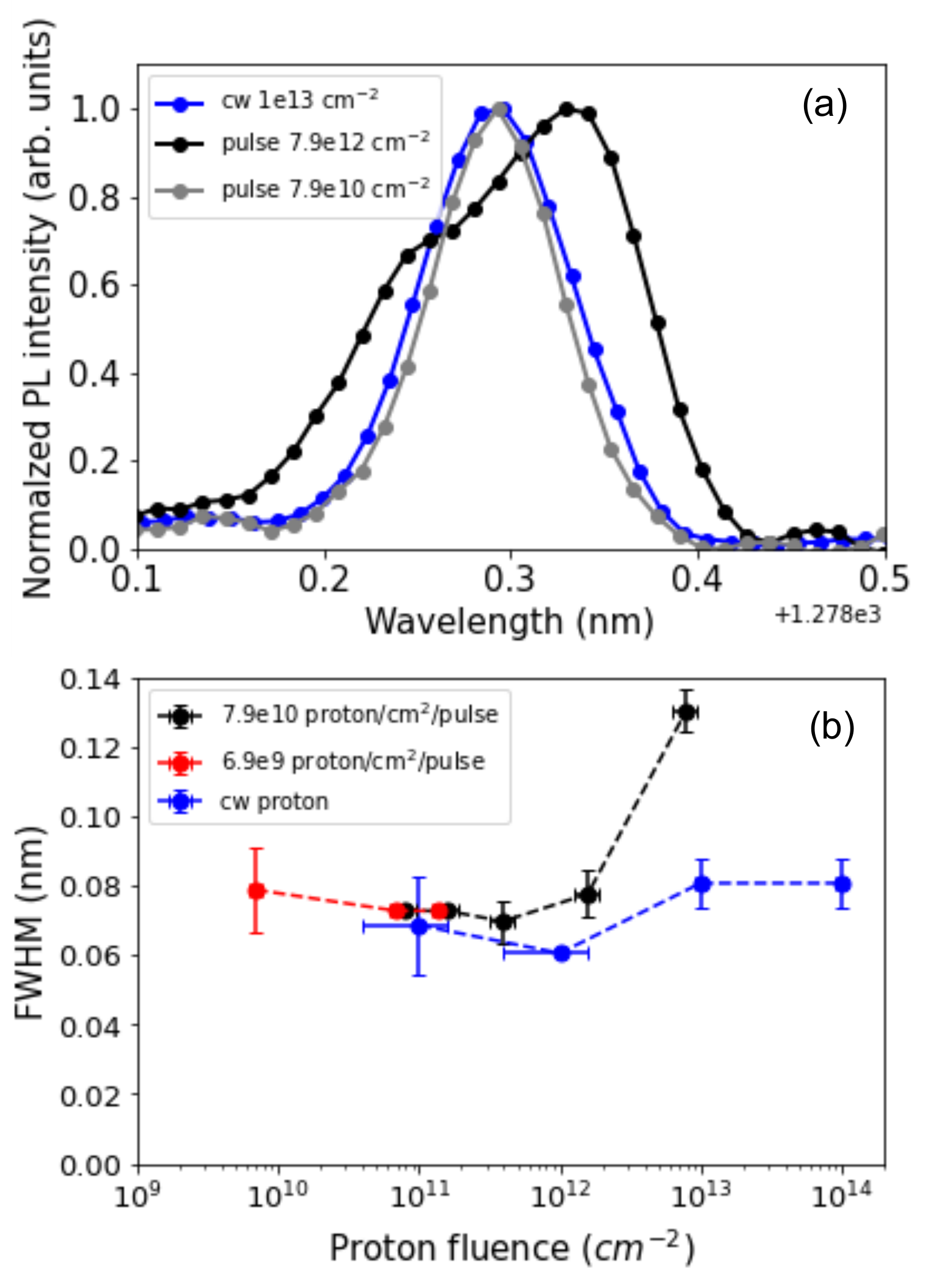}
\caption{(a) comparison of normalized ZPL spectrum of G-center generated by 100 shots (black) and 1 shot (gray) of $7.9 \pm 1.6 \times 10^{10}$protons/cm$^{2}$/pulse and cw $1 \times 10^{13}$protons/cm$^{2}$; (b) comparison of the proton fluence dependent linewidth (FWHM) of G-centers generated by $7.9 \pm 1.6 \times 10^{10}$protons/cm$^{2}$/pulse, $6.9 \pm 1.6 \times 10^{10}$protons/cm$^{2}$/pulse, and cw protons.}
\label{fig4}
\end{figure}

The optical spectral linewidth of color centers characterizes the local inhomogeneities, which are usually the dominating factor causing dephasing of a color center qubit, and are an important metric for spin-photon interface applications. As shown in figure \ref{fig4} (a), 1 shot (gray) of $7.9 \pm 1.6 \times 10^{10}$ protons/cm$^{2}$/pulse allows generating G-centers with narrow linewidths $<$ 0.08 nm, whereas the spectrum becomes significantly broadened to 0.13 nm by 100 shots of proton pulses (black). In particular, compared to the single Lorentzian-shape spectrum of 1 shot of proton pulse, the spectral broadening induced by 100 shots is comprised of two groups of G-center emission with $\pm$ 0.05 nm shift, which likely results from the local strain variation caused by the accumulation of intersitital/vacancy clusters. These subtle optical features are consistent with the high flux induced heat spike phenomenon, which leads to a super-linear damage accumulation that creates an underdense region of vacancies in the hot cascade core and an overdense region of Si interstitials pushed outward to the periphery of the cascade core \cite{ nordlund2018primary, niu2023machine, gimeno2019deterministic}. On the other hand, with a similar fluence to the 100 shot case, G-centers formed by 10$^{13}$ cm$^{-2}$ cw irradiation preserve their narrow linewidth $<$ 0.08 nm (blue in figure \ref{fig4} (a)), which is as narrow as the spectrum produced by the 1 shot pulse. It indicates that at low flux levels and for light ions, the radiation-induced intersititals and vacancies mainly recombine before damage overlap occurs. It is indeed reported that the lattice strain associated with the production of a Frenkel pair can induce spontaneous recombination of the vacancy and self-interstitial if they are located within several lattice periods from each other \cite{niu2023machine, nordlund2018primary}, which consequently reduces the local strain variation and suppresses the linewidth broadening of G-center. 

Figure \ref{fig4} (b) summarizes the variation of the G-center linewidth under cw irradiation, which remains $<$ 0.08 nm as the proton fluence increased from 10$^{11}$ cm$^{-2}$ to 10$^{14}$ cm$^{-2}$. Such a narrow linewidth is resilient to environmental charge fluctuations within an optical pump power range, as detailed in Appendix \ref{appendix:a}. Meanwhile, for the pulsed condition, the transition of linewidth broadening from 0.08 nm to 0.13 nm occurs at a fluence around 10$^{12}$ cm$^{-2}$ (which is nevertheless still narrower than the linewidth of G-centers formed by carbon and proton co-implantation with annealing \cite{beaufils2018optical,berhanuddin2012co}). It is worth noting that the PL decay of G-centers formed by the 100 shots of pulse proton is faster than that of the 10$^{13}$ cm$^{-2}$ cw irradiation, indicating that more vacancy-related nonradiative defects are being introduced by the higher dose rate. This is in line with the onset of clustering of vacancies causing a stronger degree of atomic disorder. In addition, we performed 70 keV Ar irradiation with 10$^{12}$ cm$^{-2}$ fluence on the same Si wafer, which generates damage events at a rate 3 orders of magnitude higher than 1 MeV protons with 10$^{13}$ cm$^{-2}$ fluence. We observed G-centers with broader linewidth (0.16 nm) formed by the Ar irradiation, which further suggest the formation of vacancy clusters induced by dense damage cascades. From the application aspect, our results demonstrate that the optical signal from (single) G-centers enables atomic scale sensing of radiation damage resulting from ultra-dilute, low fluence irradiations, e. g. expanding capabilities for dark matter searches \cite{ebadi2022directional}. 

\section{\textit{ab initio} Modeling of Inhomogeneous Broadening}
To understand the origin of inhomogeneous broadening in the G-center spectrum in the presence of radiation damage, first principles electronic structure calculations of the zero-phonon line (ZPL) were performed, using the approach detailed in Appendix \ref{appendix:b}. The radiation damage to the silicon unit cell hosting the G-center was simulated by introducing vacancies (Vac) at lattice sites or silicon interstitials (Si$_{(i)}$) at tetrahedral voids within the structure, in order to modify the G-center. For the type-B configuration of the G-center (GCB) embedded into a $3\times 3\times 3$ supercell of silicon, there are 215 possible silicon sites which can be replaced by vacancies including the Si$_{(i)}$ which is part of the G-center itself \cite{GC1, ivanov2022effect}, as well as 106 tetrahedral voids into which an additional Si$_{(i)}$ can be introduced. It was previously established that self-interstitials and vacancies in silicon can exist in various charge states depending on the position of the Fermi level, ranging from $-2$ to $+2$ for Si$_{(i)}$ \cite{sii_charge} and $0$,$+1$,$+2$ for Vac \cite{vac_charge}. For our configuration, we computed formation energies \cite{spinney} by setting the chemical potential at the valence band maximum, to establish the $+2$ as the most stable charge state for both defects, consistent with prior studies. We note that the dimension of the silicon supercell (1.63 nm)$^3$ and periodic boundary conditions limit the maximum separation between GCB and defect to about $\sim$1 nm, which is considerably less than the average separation generated by experiments. Nevertheless, the modeling provides a qualitative understanding of the broadening and shape of the emission spectrum. In addition, we note that some of these configurations were excluded from the plot, because of nonconvergence within the density functional theory self-consistent cycle as a result of the close proximity of GCB and interstitial/vacancy.

\begin{figure}[ht]
\includegraphics[width=\columnwidth]{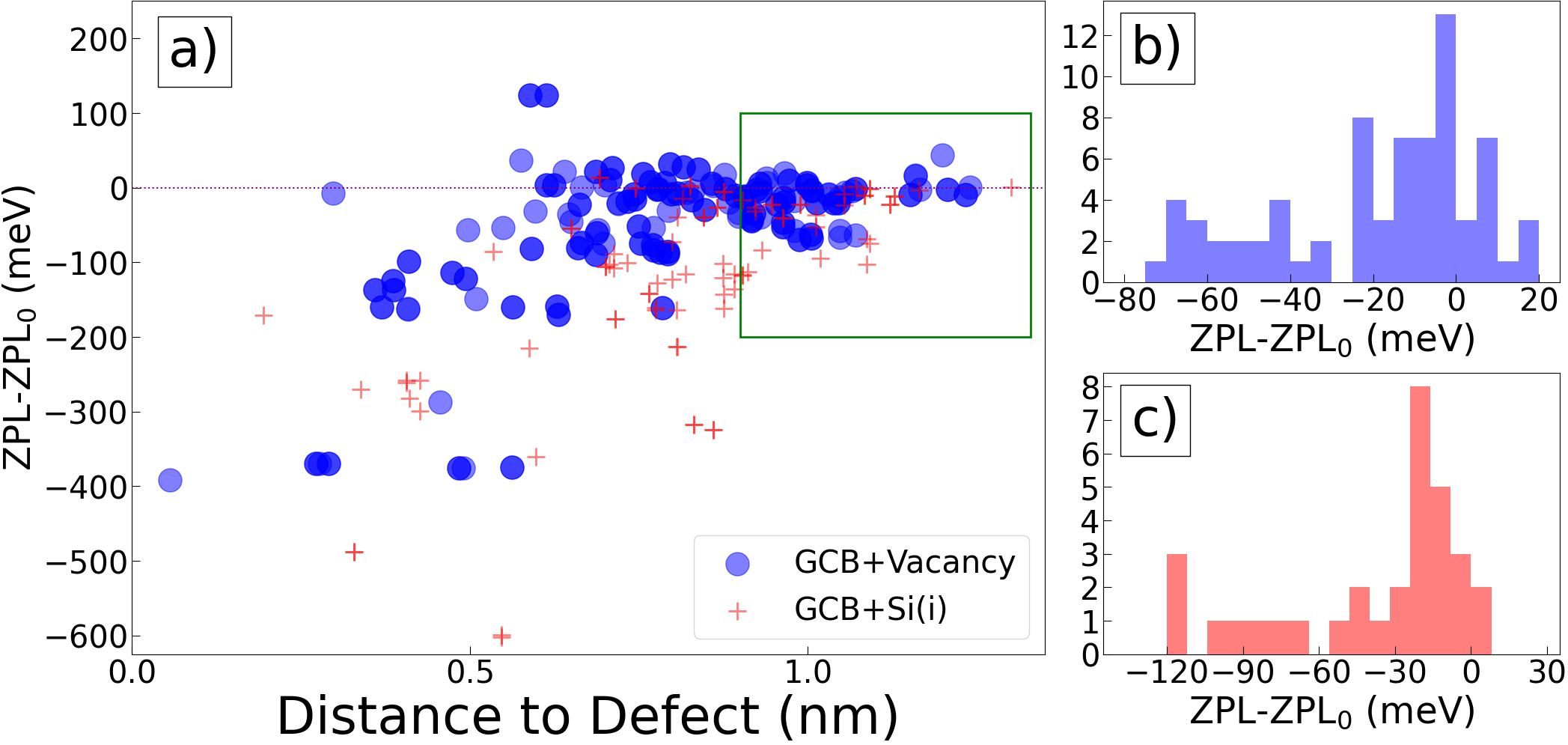}
\caption{Shift of ZPL transition energy (a) vs distance between GCB and the perturbing defect, where the defect is a vacancy (blue circles), or silicon self-interstitial (red cross). Histograms of the ZPL shift due to vacancies (b), and self-interstitials (c) for the region marked by the green box in (a).
}
\label{fig5}
\end{figure}

For these 321 modified G-center structures, force-relaxed ground and excited state structures were computed to obtain the ZPL transition energies, shown in Figure \ref{fig5} (a). The distribution of ZPLs is broadened dramatically along with a collective redshift when defects are closer than 0.5 nm to the G-center. The behavior of configurations where the defects are farthest ($\sim 1$ nm) from GCB has previously been identified as relatively closely representative of experiment \cite{redjem2022defect}. For separations $>0.9$ nm, the distribution of ZPLs is considerably narrower, and we plot histograms of the ZPL shifts for Vac and Si$_{(i)}$ in Figures \ref{fig5} (b) and (c), respectively. These show that vacancies appear to generate bilateral red- and blue-shifts in the ZPL, while Si$_{(i)}$ have a tendency towards redshift.

\begin{figure}[ht]
\includegraphics[width=\columnwidth]{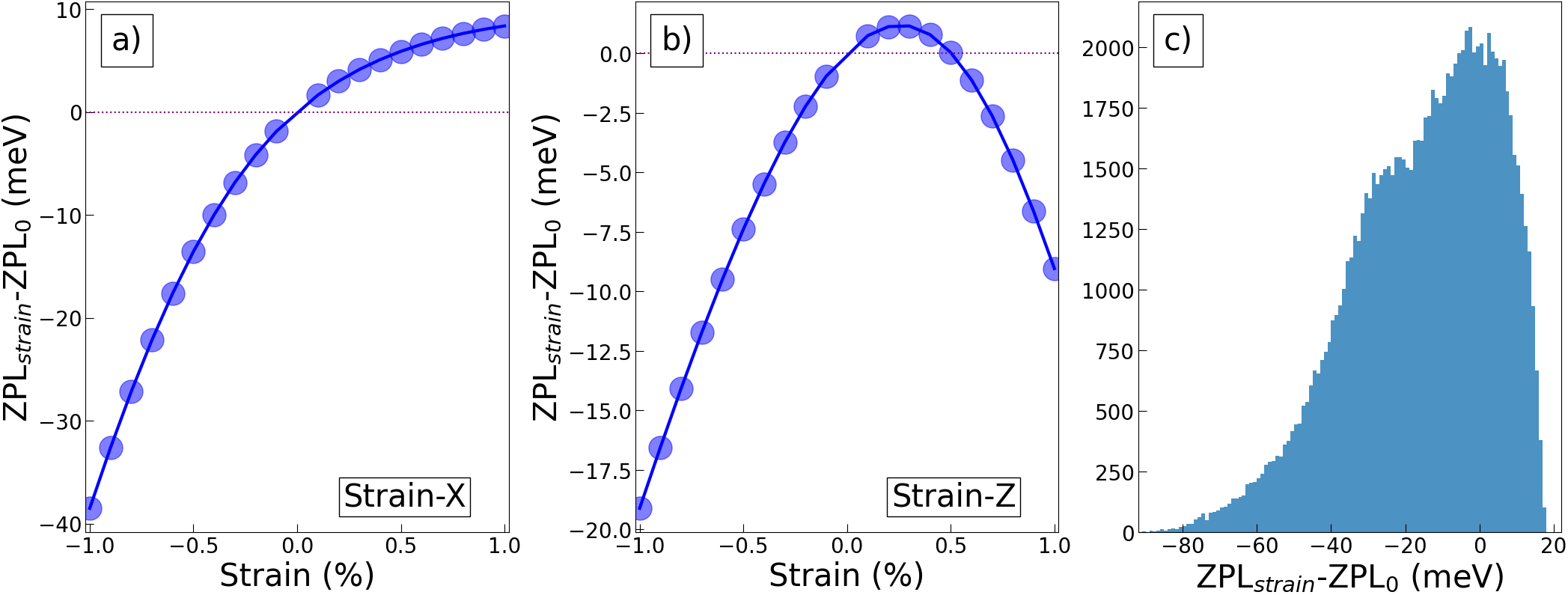}
\caption{ZPL shift vs strain for $x$ (a), and $z$ (b) directions. Distribution of combined ZPL shifts for a uniform selection of $<x,y,z>$ strain vectors.}
\label{fig6}
\end{figure}

The effect of the local defects on the ZPL of GCB can be understood in terms of the local strain they induce. To isolate the impact of local strain, we perform strain calculations on an isolated G-center ranging from -1\% to +1\% in increments of 0.1\% for $x$,$y$, and $z$ linear strains, $xy$,$xz$, and $yz$ in-plane strains, and isotropic $xyz$-strain \cite{supplement}. In addition, due to the $<110>$ mirror plane symmetry of the G-center, strains along the $x$ and $y$ directions will be equivalent, and strains along the $xz$ and $yz$ will be equivalent as well. The ZPL shifts due to strain in the $x$ and $z$ directions are shown in Figure \ref{fig6} (a) and (b), and can be used to interpret the effect of vacancies and silicon self-interstitials on the G-center emission. In particular, a vacancy will result in nearby silicon atoms relaxing to fill the void and lead to a tensile strain that expands the lattice around any nearby G-centers, as was proposed previously \cite{redjem2023all}. This tensile strain can result in either a redshift or blueshift of the ZPL depending on the relative position of the vacancy and GCB, explaining the bimodal distribution seen in Fig. \ref{fig5}b. Conversely, Si$_{(i)}$ would lead to compressive strain on GCB, which always results in a redshift (Fig. \ref{fig5}c), regardless of the defect position.

More detailed modulation of the distribution of ZPL shifts occurs when we further consider a distribution of defect locations (strains). As an example, we select a uniform distribution of $10^5$ strain vectors with random $x$,$y$,$z$ components ranging from -1\% to +1\% , and interpolate the effect on the ZPL by adding the shift generated by each strain direction.
The resulting ZPL distribution is shown in Fig. \ref{fig6} (c), 
and is consistent with the histogram plots in Figures \ref{fig5}(b)and (c), showing the characteristic redshifted shoulder. For this distribution of strains, the resulting ZPL distribution may be explained by $x$-directional strains favoring large redshifts, and $z$-directional strains causing comparatively smaller ZPL shifts. We also note that the proton irradiation will not necessarily result in a uniform distribution of vacancies/interstitials around the G-centers. In Appendix \ref{appendix:b}, we compute the distribution of combined ZPL shifts for a sample of $<x,y,z>$ strain vectors heavily biased in the $z$-direction. 

While these first-principles calculations were performed for small GCB-defect separations, their interpretation in terms of strain fields allows us to understand the behavior at larger GCB-defect separations present in our samples. A possible explanation for the two peaks in the experimental PL spectra (Fig.~\ref{fig3}) is the presence of two separate populations of defect-induced strain: one arising from interstitials resulting in redshifts, and one from vacancies, resulting in blue- or redshifts clustered around a different mean value. Based on our predictions of the ZPL-strain relation, we estimate the experimental local strains around the G-centers to be of the order 0.01 \%.
Therefore, the collective strain introduced by decorating vacancies and interstitials around the G-center, qualitatively explains the experimentally observed spectral broadening and dominant redshift of the G-center spectrum caused by high flux and high fluence pulsed proton irradiation in figure \ref{fig4} (a).

\section{Conclusions}

In conclusion, we have studied the formation dynamics of G color centers in silicon far from equilibrium under various dose rates of proton beams. Compared to continuous wave proton irradiation, pulse excitation offers the benefit of enhanced transient excitations and lattice heating by the instantaneous energy deposition in about 10 ns long proton pulses, which allows optimizing the ratio of G-center formation to destruction, and to nonradiative defect accumulation within the nanosecond time regime. By characterizing the PL of G-centers, our results highlight the approach of pulsed proton irradiations on silicon wafers  to generate color center complexes with high formation efficiency while preserving narrow linewidths. Moreover, compared to the broadening of G-center ensemble linewidths by residual lattice damage from carbon ion implantation and thermal annealing, the narrow linewidths of G-centers created by our approach allows them to be used as a sensitive probe of atomic radiation damage when the irradiation condition is above a certain threshold of dense damage cascades. PL of G-centers is a readily available analytical method for radiation damage studies \cite{davies1989optical}. Aided by \textit{ab initio} electronic structure calculations, we provided insight into the atomic disorder-induced inhomogeneous broadening by introducing vacancies and Si interstitials in the vicinity of a G-center. A vacancy leads to a tensile strain can result in either a redshift or blueshift of the G-center emission, depending on its position relative to the G-center. Meanwhile, Si interstitials lead to compressive strain, which results in a monotonic redshift. The intense and tunable proton and ion pulses (from induction accelrators or laser-acceleration) enable enable studies of fundamental defect dynamics of radiation-induced defects, defect engineering, and qubit synthesis for quantum information processing.
\nocite{*}

\begin{acknowledgments}
Work at Berkeley Lab was supported by the Office of Science, Office of Fusion Energy
Sciences, of the U.S. Department of Energy, under Contract No. DE-AC02-05CH11231 and
by the Molecular Foundry, a DOE Office of Science User Facility supported by the Office of
Science of the U.S. Department of Energy under Contract No. DE-AC02-05CH11231. This research used resources of the National Energy Research Scientific Computing Center,
a DOE Office of Science User Facility supported by the Office of Science of the U.S.
Department of Energy under Contract No. DE-AC02-05CH11231.
\end{acknowledgments}

\appendix
\section{Pump power dependent optical properties}
\label{appendix:a}
To characterize the impact of optical excitation on the G-center optical properties, we measured the PL signal from a sample irradiated by 20 shots of $6.9 \pm 1.6 \times 10^{9}$ protons/cm$^{2}$/pulse by varying the laser power from 0.3 mW to 1.2 mW (before the objective).
\begin{figure}[ht]
\includegraphics[width=\columnwidth]{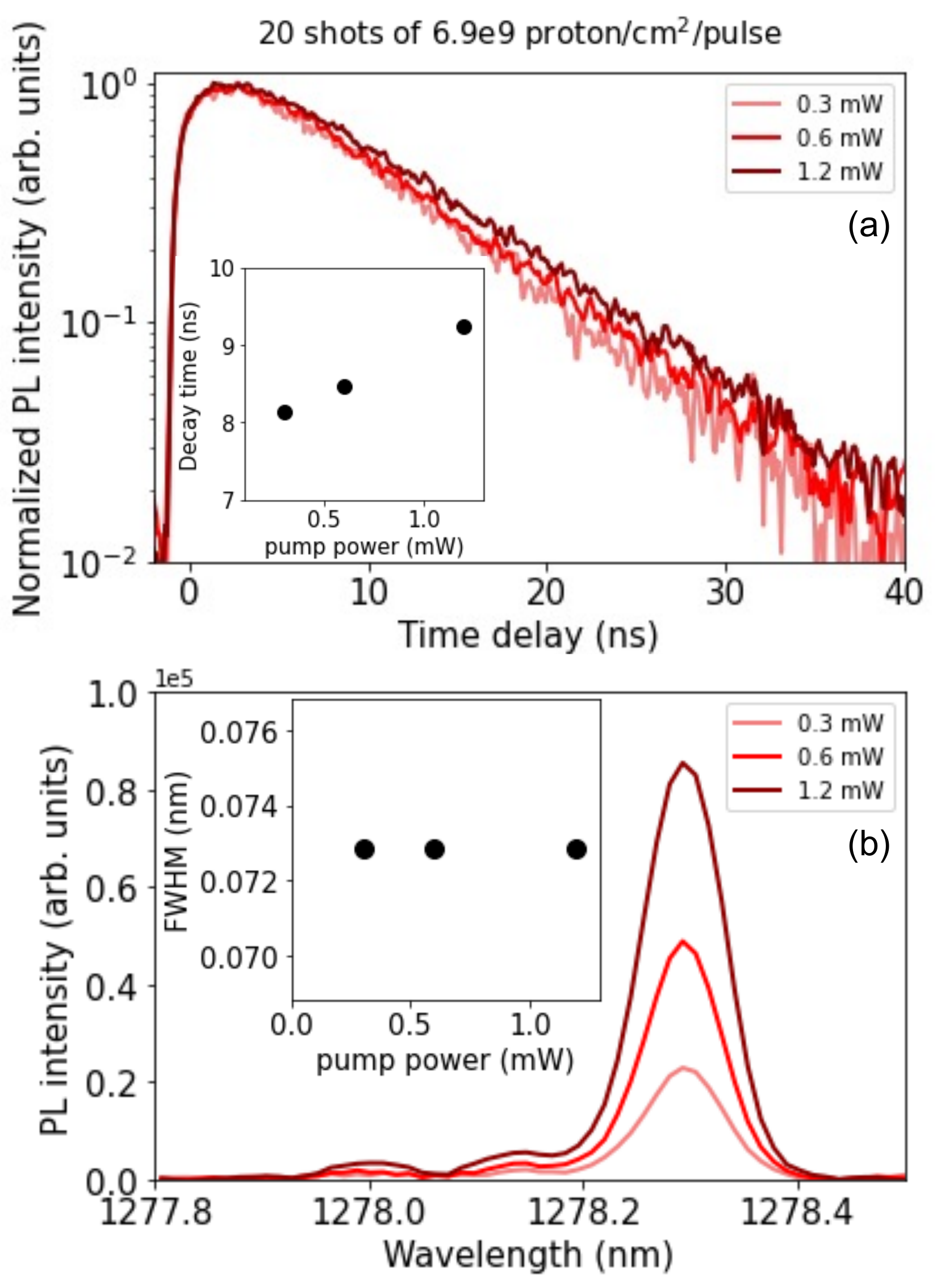}
\caption{\label{fig:wide}(a) the PL decay of G-center emission with the inset showing the PL decay time extracted by single exponent fit; (b)the corresponding pump power-dependent PL spectra of the G-center, and the FWHM of the PL spectrum (inset).}
\label{fig3}
\end{figure} 
As shown in figure \ref{fig3} (a), when the pump power is increased, the PL decay of G-center emission becomes slower, with the decay time increasing from 8.1 ns to 9.3 ns (see the inset of PL decay time extracted by single exponent fit). Such a trend corresponds to the typical feature of gradual saturation of the nonradiative SRH process. Figure \ref{fig3}(b) shows the corresponding pump power-dependent PL spectra of G-center, with main G-center peak at 1278.3 nm formed by naturally abundant $^{28}$Si. Additionally, the spectra also shows two blue-shifted side peaks, which are associated with the isotope shifts of $^{29}$Si and $^{30}$Si occupying the interstitial site, respectively \cite{chartrand2018highly}. The inset in \ref{fig3}(b) plots the FWHM of main ZPL at 1278.3 nm, which keeps constant and as narrow as 0.073 nm. It indicates the G-center linewidth is insensitive the environmental charge fluctuation in this pump power range.

\section{Modeling of ZPL distribution with strain vectors biased in z-direction}
\label{appendix:b}

The first principle electronic structure calculations used the Vienna \textit{ab initio} simulation package (VASP) \cite{vasp1,vasp2,vasp3,vasp4}, employing an HSE06 hybrid-functional approach \cite{hse,Gcenter-HSE} to model the silicon G-center in the B-type (GCB) configuration \cite{GC1, gali-GC}. The silicon G-center was embedded into a $3\times 3\times 3$ supercell of silicon containing 216 silicon atoms, and the forces were relaxed to a tolerance of 0.001 (eV/\AA) on a $1\times 1 \times 1$ $\Gamma$-centered $k$-point grid, with an energy cutoff of 450 eV. To obtain ZPL values, excited state calculations were performed using the constrained occupation method \cite{dhaliah2022first,gali-GC, ivanov2022effect}, with a full force relaxation step to the same tolerance noted above.

\begin{figure}[h]
\includegraphics[width=\columnwidth]{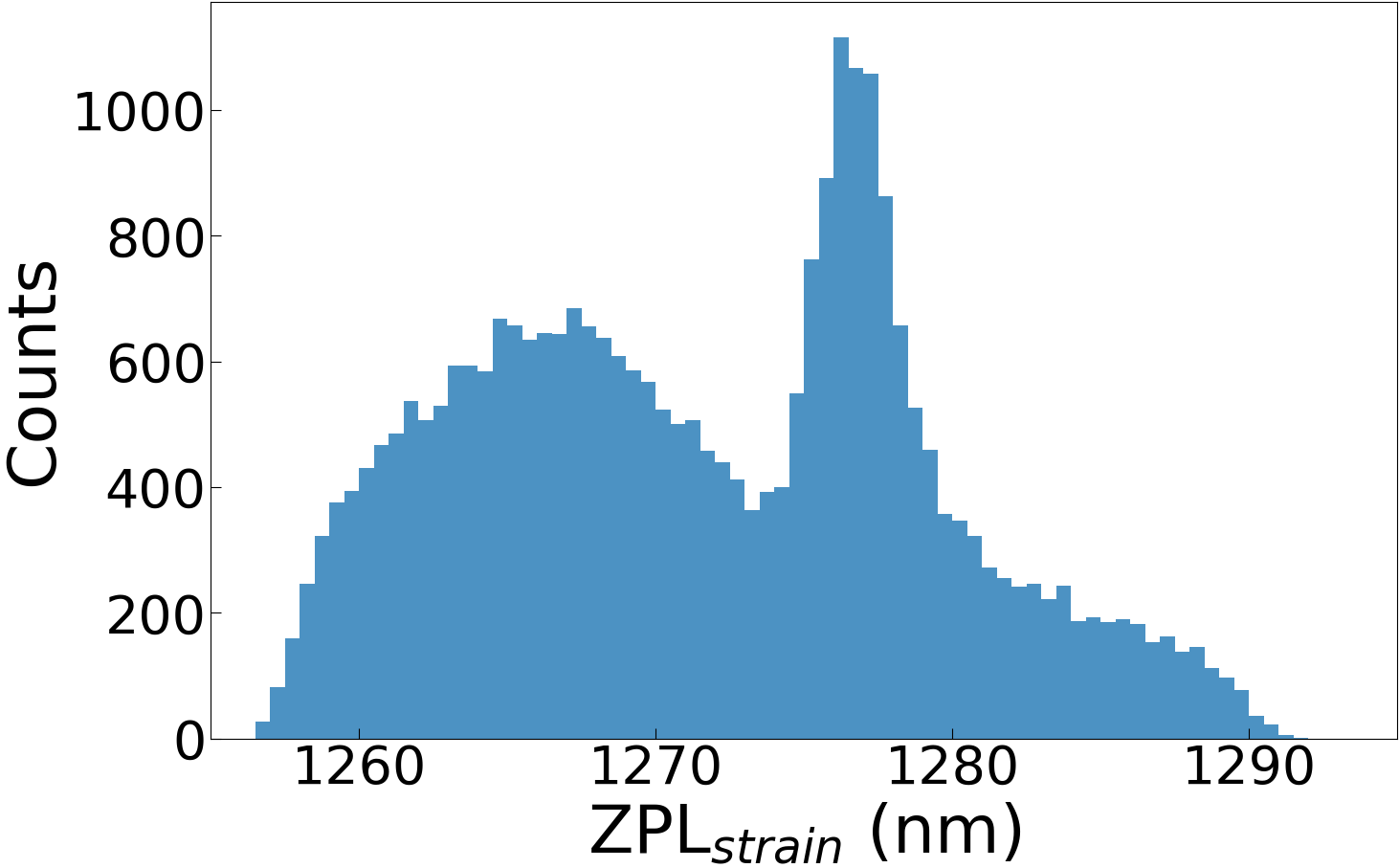}
\caption{Distribution of combined ZPL shifts for a sample of $<x,y,z>$ strain vectors heavily biased in the $z$-direction.}
\label{fig7}
\end{figure}

To model a distribution of vacancies/interstitials oriented along the $z$-direction, we repeat the procedure for a uniform distribution described in the main text, but discard at random all but $1/10$ of the strain vectors that exceed 0.1\% in the $xy$ directions. The ZPL shifts generated in this way create the spectrum shown in Fig.~\ref{fig7} with a broad blueshifted component and a relatively concentrated component near the original ZPL, which is qualitatively similar to our previous experimental spectrum obtained in the high-damage limit \cite{redjem2022defect}, where such a biased distribution of local strains might be expected. In principle, such a procedure can be inverted to estimate possible local strain distributions from measured spectra, and in turn deterimine the distribution of defect damage near silicon G-centers. 

\bibliography{reference}
\end{document}